\documentstyle{article}
\newcommand{\beq}{\begin{equation}}
\newcommand{\eeq}{\end{equation}}
\newcommand{\bara}{\begin{eqnarray}}
\newcommand{\ear}{\end{eqnarray}}
\newcommand{\hp}{$|\kern-0.2em \rightarrow \kern-1em \leftarrow \rangle
$  }
\newcommand{\vp}{$|\kern-0.2em \downarrow \kern-0.5em \uparrow \rangle $
}
\newcommand{\rp}{$|\kern-0.2em \nearrow \kern-1em \swarrow \rangle $  }
\newcommand{\lp}{$|\kern-0.2em \nwarrow \kern-1em \searrow \rangle $  }

\input{epsf}
\begin{document}

\title{``Plug and play'' systems for quantum cryptography}
\author{ A. Muller \and T. Herzog \and B. Huttner \and W. Tittel \and H.
Zbinden \and N. Gisin\\
Group of Applied Physics\\University of Geneva\\
20 Rue de l'Ecole de Medecine\\
CH 1211 Geneva 4\\ Switzerland}

\maketitle

\begin{abstract}
We present a time-multiplexed interferometer based on Faraday mirrors, and
apply it to quantum key distribution. The interfering pulses follow
exactly the same spatial path, ensuring very high stability and self
balancing. Use of Faraday mirrors compensates automatically any
birefringence effects and polarization dependent losses in the
transmitting fiber. First experimental results show a
fringe visibility of 0.9984 for a 23km-long interferometer, based on
installed telecom fibers. 
\end{abstract}


In so-called private-key cryptographic systems, secure transmission
through unprotected channels rely on the exchange of secret keys, known
only to the sender, Alice, and the receiver, Bob. Quantum cryptography
(QC) relies on the properties of quantum mechanics to obtain a provably
secure key distribution~\cite{bennett92,bennett92jc,hughes95cp,jmo}. Most
exisiting implementations
rely on either the
polarization~\cite{bennett92jc,mullerqc93,muller95} or the 
phase~\cite{bennett92a,townsend93a,townsend94a,hughes95rc} of very weak
pulses of light as information
carrier. To date, the longest transmission spans were obtained in optical
fibers, at a wavelength of 1300 nm~\cite{muller95,townsend94a,hughes95rc}. 

The main difficulty with polarization-based systems is the need to keep
stable polarizations over distances of tens of kilometers, in standard
telecom cables. Indeed, due to the birefringence of the fibers and the
effect of the environment, the output polarization fluctuates randomly.
Recent experiments~\cite{muller95} have shown that in general the
time-scale of these
fluctuations is long enough (tens of minutes) to enable polarization
tracking to compensate for them. However, while this is no major problem
for preliminary experiments, this would be inconvenient for practical
applications of QC. 

\begin{figure}[p]
\hspace*{1cm}\epsfbox{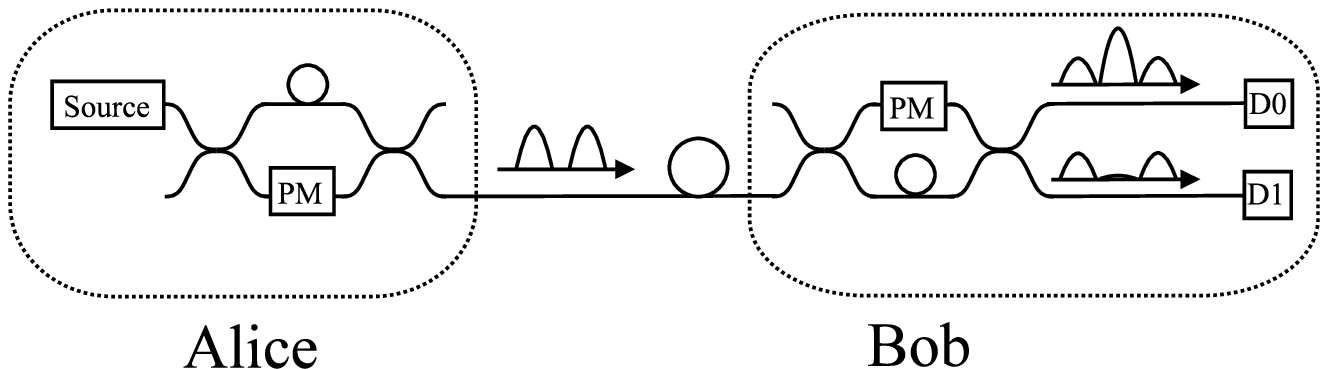}
\caption{Mach-Zehnder interferometer for QC.}
The pulsed laser is split into two time-shifted pulses by Alice: one goes
trough the short path (S) and the phase modulator (PM); and the second
through the long path (L). Information about the key is encoded in the
phase-shift introduced in PM. After propagation, the two pulses arrive
into a similar interferometer on Bob's side, creating three pulses. The
first one (SS) and the last one (LL) carry no information on the phase
settings. The middle one is obtained by interference between two paths: SL
and  LS. The relative phase settings creates a constructive or destructive
interference in D0 and D1. In order to obtain a good visibility, the two
interferometers have to be kept identical and should preserve
polarization.  
\label{MZ}
\end{figure}

Interferometric quantum key distribution systems are usually based on a
double Mach-Zehnder interferometer~\cite{townsend94a,hughes95rc}, one side
for Alice and one for Bob
(see Fig.~\ref{MZ}).  These interferometers already implement
time-multiplexing, as both interfering pulses follow the same path between
Alice and Bob, with some time delay.  However, the pulses do follow
different paths within both Alice's and Bob's interferometers. In order to
obtain a good interference, both users therefore
need to have identical interferometers, with the same coupling ratios in
each arm and the same path lengths, and also need to keep them
stable within a few tens of nm during a
transmission.
Moreover, since optical components like phase modulators (PM) are
polarization dependent, polarization control is still necessary both in
the transmission line and within each
interferometer. This
again is inconvenient for practical applications. 

In this letter, we present a new interferometric system implementing
phase-encoded quantum key distribution~\cite{patent}.
It is based on time-multiplexing, the interfering pulses now following
exactly the
same spatial path, albeit with a small time delay. Therefore, in contrast
to the usual schemes, it does not require any path length control between
the various paths. Moreover, all pulses are reflected back at the end of
the fibers. Use of Faraday mirrors instead of regular mirrors makes it
possible
to suppress all birefringence effects and polarization dependent losses
occuring during the transmission.
Therefore, our system does not require any polarization control.  In
essence, with our system, Alice and Bob could exchange their cryptographic
keys through standard telecom systems, with no need for lengthy
adjustments. They would be provided with a sending kit and  a
receiving kit, and could simply plug them in at the end of the fiber,
synchronize their signals, and start the exchange. This is the reason why
we informally refer to our system as a ``plug and play'' system. 

\begin{figure}[p]
\hspace*{1cm}\epsfbox{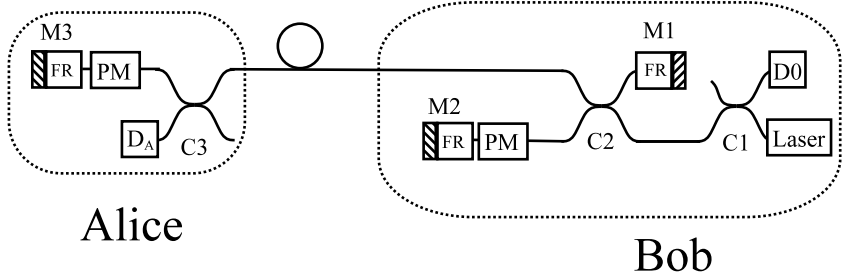}
\caption{``Plug and play'' system with phase-encoding.}
A short laser pulse sent by Bob is split into two at coupler C2. The first
part, P1 goes straight to Alice, while the second one, P2, is first
delayed by
the M2-M1 delay line. Both pulses are reflected back towards Bob at M3.
Alice measures the intensity of the incoming pulses,
and attenuates them to single-photon levels. The phase modulators (PM)
modulate the path length between the two pulses. On arrival to Bob's side,
part of P1 is delayed by M1-M2, and thus interferes with incoming P2. The
interference pattern at D0 gives the relative phase settings of Alice and
Bob. Use of the Faraday rotators (FR) before the mirrors makes it possible
to cancel
out all birefringence effects in the fibers. 
\label{phase-set}
\end{figure}

A schematic of the setup is given in Fig.~\ref{phase-set}. Bob initiates
the transmission by sending a short laser pulse towards Alice. Let us
for the moment disregard the effects of the Faraday rotators (FR).  The
need for coupler C3 and detector $ \rm D_A$ in
Alice's
arm will also be explained later. 
The pulse arriving in C2 is split into two parts: one part, P1, goes
directly towards Alice; while the second part, P2, is first delayed by one
bounce in the M2-M1 delay line. The two pulses, P1 and P2,
travel down the fiber to Alice. In order to encode her bits, Alice lets P1 
be reflected by M3, but modulates the phase of P2.  Detection
on Bob's side is done by delaying part of P1 in the same M1-M2 delay line
(using also another PM, this time on P1), and looking at the interference
with P2. 
If the PMs at both Alice's and Bob's are off, the interference is
constructive (the two pulses follow exactly the same path). 
If however Alice or Bob change
their phase setting betwen the two pulses, the interference may become
destructive. In fact it is easily seen that destructive interference
is obtained when: $ \phi_{\rm A} - \phi_{\rm B}= \pi$, where $
\phi_{\rm
A}$ and $\phi_{\rm B}$ are the total phase  shifts introduced by Alice and
Bob
respectively. In this case no light is detected at D0. This shows that
the relative phase setup modulates the intensity in D0, and thus can be
used to transfer information from Alice to Bob. The first attractive
features of this setup are that the interferometer is automatically
aligned (both pulses are delayed by the same delay line), and that the
visibility of
the fringes is independent of the transmission/reflection
coefficients of C2. 
 Of course, a large fraction of the light does not follow
these two paths, but is split differently at the various couplers (e.g.
keeps oscillating a few times between
M1-M2 or M1-M3 before leaving towards D0). These pulses will
eventually arrive in D0, but at a different time, and will be easily
discriminated. Therefore, they do not reduce the visibility. 

The above setup would work perfectly well for ideal fibers, with no
birefringence. Unfortunately, all existing optical fibers have
birefringence and polarization couplings, which will modify randomly the
state of polarization of the light,
and may lead to a reduction in the visibility of the
interference. 
In order to preserve
interference, we replace the mirrors by so-called Faraday
mirrors (FM)~\cite{martinelli92}. A Faraday mirror is  simply an ordinary
mirror, glued on a
Faraday rotator (FR), which rotates the polarization by $45^{\circ}$. The
effect of a FM is to transform any polarization state into its orthogonal.
The most interesting consequence is that a FM automatically compensates
any
birefringence effect in the fiber:  the  state going out of the fiber is
alway orthogonal
to the incoming state.  Replacing
the ordinary mirrors M1 and
M2 by FMs (i.e. adding
the FRs), thus ensures that the two pulses P1 and P2 have the same
polarization, irrespective of birefringence effects in the delay line
M1-M2. Use of an extra  FM in M3 enables to compensate for the
polarization dependence of the PM. 

Till here, we have only discussed macroscopic pulses. In order to get
quantum cryptographic security, the information carrying pulses need to be
very weak, with
at most one photon per pulse. This is to prevent a malevolent
eavesdropper, known as Eve, to divert part of the pulse and get
information on the key (see e.g.~\cite{bennett92jc} for more on
eavesdropping). In practice, we rely on strongly attenuated laser light,
with about 0.1 photon per pulse on average. This attenuation is obtained
by adding the extra coupler C3 in Alice's arm.
Using detector $ \rm D_A$, Alice can monitor the intensity of the incoming
pulses, and attenuate them to ensure that P2 going back to Bob has indeed
the correct intensity (remember that
the pulses going from Bob to Alice do not carry any phase information yet;
it is only on the way back to Bob that the phase chosen by Alice is
encoded in P2). 
In order to be able to
use ordinary detectors, and not single-photon ones, it is preferable for
Alice to have a strongly transmitting coupler, with $t_3 \approx 1$. This
maximizes the intensity going to her
detector, as well creating enough attenuation on the beams reflected by M3
to have a
single-photon-like pulse sent back to Bob. Monitoring the
incoming intensity has the added advantage that Alice can  detect any
attempt
by Eve to obtain the value of her
phase shift by sending much stronger pulses in the system, and measuring
the phase of the reflected pulses. On Bob's side, detector D0 needs to be
a single-photon detector. In order to obtain as much of the light  as
possible on D0, the coupler C1 has to be  mostly transmitting.

Let us now show how this scheme implements the original phase-encoding
proposal~\cite{bennett92a}, known also as the two-states system. Both
Alice and Bob choose at random their phase settings: 0 or
$\pi$ phase shifts, corresponding respectively to bit value 0 and
1.  If Alice
and Bob use different phase shifts, the difference is always $\pi$,
which means that the interference in D0 is always destructive. So, if Bob
chooses bit 0, and gets one count in his detector, he knows that Alice has
also sent a 0, and reciprocally for bit 1.  Of course, since
they use very weak pulses, in many instances Bob would get no count in D0.
In this case, he cannot infer what was sent by Alice: it could be
that  Alice used a different phase; or it could be that there was
simply no photon in the pulse.  This corresponds to the fact that the two
states are not orthogonal, and thus cannot be distinguished with
certainty~\cite{bennett92a}.  We can now understand why we do need very
weak pulses: if Alice and Bob use strong pulses, which always carry more
than one photon, Bob would always know the bit sent by Alice: one count,
same choice of phase; no count, different choice of phase.  Unfortunately,
so would Eve. For
example, she could simply add an extra coupler on the line, and measure
the phase of the pulses sent by Alice. However, if the pulse sent by Alice
possesses at most one photon, this simple eavesdropping strategy fails
completely: if Eve measures the photon, then Bob will not get it, and
would simply discard the corresponding transmission. 

Another eavesdropping strategy on two-state systems, discussed
in~\cite{bennett92a}, would be
for Eve to stop the transmission altogether, measure as many pulses as she
could, and send to Bob only the ones for which she managed to obtain the
phase. To prevent
this, Alice needs to send both a strong pulse, as a reference,  and a weak
one, containing the phase information. Eve cannot suppress the strong
pulse without being immediately discovered. If she supresses only the weak
one, because she did not obtain the phase information, the strong pulse
alone will introduce noise in detector D0. In our system, this is easily
implemented by using a strongly asymmetric coupler C2, with transmission
coefficient $t_2 \approx 1$.  In this case, P1 going back towards Bob is
much stronger than P2, which has
already been through the M1-M2 delay line, and thus was strongly
attenuated
(by a factor $(r_2)^4$ in the intensity). Bob
can detect the part of P1 going directly to detector D0, before looking at
the interference. It is also possible to simply add an extra coupler and
detector in front of M1, in a
way similar to Alice's setup. Further discussion on eavesdropping would be
outside the scope of this paper~\cite{ekert94}. 

The same setup, but with different choices of phase for Alice and Bob can
trivially be used to implement other protocols, such as the well-known
BB84 protocol~\cite{bennett92jc}, where Alice chooses between four
possible states. A very similar one can also be used for
polarization-encoded systems. It suffices to replace the coupler C2 by a
polarization coupler (PC2), and send the light with circular polarization.
One of the polarization components, say the vertical one, follows the path
of P1 (with a polarization switch from vertical to horizontal and vice
versa each time it is reflected by a FM), while the horizontal one follows
the path of P2. A phase change on Alice's side now corresponds to a
different output polarization.
This system does not require a second PM on Bob's side, but needs a more
complicated detection system, which can separate the various
polarizations. 
Experimentally, we concentrated  on the phase-encoded one, which has
the
simplest detection setup. 

Our experiment implements a slightly simplified version of the
phase-encoding system. The long arm of the interferometer, which
corresponds to the distance between Alice and Bob, is one optical fiber in
a 23km long commercial optical cable, used for telecommunications between
Nyon and Geneva in Switzerland. The laser produces 300ps long pulses at
1300nm, with a repetition rate of 1kHz. The phase modulator on Bob's
side was built with a fiber wrapped around a piezoelectric modulator. This
setup has very low losses, but is limited to a few kHz. The length of the
M1-M2 delay line is 23m, which corresponds to a time delay of 250ns
between the two pulses. On  Alice's side, the phase modulator  is a
Lithium Niobate waveguide, that can be driven up to a GHz. This fast
modulation is needed in order to be able to modulate the phase of the
second pulse arriving at Alice's, without changing the phase of the first
one. One interesting
point is that this type of phase modulator is always polarization
dependent. However, this polarization dependence is cancelled out, thanks
to the Faraday mirrors (the light goes in the PM at some polarization, and
again after reflection at M3 with the opposite one). At this stage, we
do not insert  detector $D_A$, but only modulate the phase of the pulse.
The visibility of the
interference fringes is 0.9984, without any need for adjustments. The
system is also totally stable. 

During the mere few years since the first experimental demonstration of
QC, we have seen tremendous progress. The first setup in 1992 used visible
light, over an air gap of about 30 cm. Recent results show that quantum
key distribution is possible over distances of tens of kilometers, using
only standard telecom cables. The main experimental challenge is now to
demonstrate that QC is practical. Existing systems, while perfectly fine
for a demonstration, require careful adjustements and control of the
systems on each side of the communication channel. In this letter, we
proposed what we call a   ``plug and play'' system, which requires no
adjustement at all, except for the timing. Our first experimental results
show a very good stability, and high fringe visibility.  This shows that
our new scheme is indeed very promising for practical implementations of
QC. We are currently working on a fully operational prototype.

{\bf Acknowledgements\\}
We would like to thank the Swiss PTT for financial support, and for
allowing access to the Nyon-Geneva optical cable.

\newpage

\end{document}